\documentclass[aps,prb,twocolumn,superscriptaddress,showpacs]{revtex4}
\usepackage{graphicx}
\usepackage{calc}
\usepackage{bm}

\begin{document}

\title{Separately  contacted  edge states at high imbalance in the integer and fractional quantum Hall effect regime}

\author{E.V.~Deviatov}
\email[Corresponding author. E-mail:~]{dev@issp.ac.ru}
 \affiliation{Institute of Solid State
Physics RAS, Chernogolovka, Moscow District 142432, Russia}

\author{A.~Lorke}
\affiliation{Laboratorium f\"ur Festk\"orperphysik, Universit\"at
Duisburg-Essen, Lotharstr. 1, D-47048 Duisburg, Germany}

\date{\today}

\begin{abstract}
This review presents experimental results on the inter-edge-state
transport in the quantum Hall effect, mostly obtained in the
regime of high imbalance. The application of a special geometry
makes it possible to perform $I-V$ spectroscopy between individual
edge channels in both the integer and the fractional regime. This
makes it possible to study in detail a number of  physical effects
such as the creation of topological defects in the integer quantum
Hall effect and neutral collective modes excitation in fractional
regime. The while many of the experimental findings are well
explained within established theories of the quantum Hall effects,
a number of observations give new insight into the local structure
at the sample edge, which can serve as a starting point for
further  theoretical studies.
\end{abstract}


\maketitle

\section{Introduction: edge states at low imbalance}

 Both the integer quantum Hall effect
(IQHE) and the fractional quantum Hall effect (FQHE) occur in
high-mobility two-dimensional electron systems in a quantizing
magnetic field under low temperatures. Although the  Fermi level
is within the spectrum gap in both regimes, the origins of the gap
are substantially different for the IQHE and FQHE. The integer
quantum Hall effect is explained by the Landau quantization in the
spectrum of the two-dimensional electron system in a magnetic
field. On the contrary, the FQHE is fully recognized as a
manifestation of the electron-electron interaction. Despite these
differences, charge transport is mostly determined by edge effects
in both IQHE and FQHE regimes. The present report is dedicated
 to a detailed investigation of intra-edge transport and
the differences and similarities in the physical effects observed
in both regimes.

\subsection{Edge states definition}
Halperin~\cite{halperin82} introduced current-carrying edge states
as the intersections of the Landau levels and the Fermi level near
the sample edges. Hence, the total number of edge states is equal
to the filling factor, i.e., the number of filled Landau levels,
and their electrochemical potentials are equal to the
electrochemical potentials of the corresponding edges of the
sample. If the number of filled Landau levels is $n$, the total
current through the sample can be written as $I=n (e/h)
\Delta\mu$, where $\Delta\mu$ is the difference in the
electrochemical potentials of the sample edges. Hence, the current
is determined only by the difference in electrochemical potentials
of the edges (or, in other words, of the edge states) and the
number of filled Landau levels (the number of edge states). If we
introduce the edge-state current $(e/h)\mu$, then the sum of all
edge-state currents gives the total current through the sample.

B\"uttiker~\cite{buttiker88} combined Halperin's idea of current-carrying edge states
with the Landauer formalism~\cite{landauer82}, aiming to take scattering in
one-dimensional semiconductors into account. He showed
that the effects of elastic and nonelastic scattering in edge
states and contacts can be taken into account by introducing the transmission coefficient matrix $T_{ij}$. He suggested
the formalism for calculating various resistances for
samples with many ohmic contacts. In this
formalism, the current $I_i$ carried by edge states going from a contact $i$ is given by
\begin{equation}
I_i=\frac{e}{h} \left(n_i\mu_i+\sum_{j\ne i} T_{ij}\mu_j\right)
\label{butt_form},
\end{equation}
where $I_i$  is the current through edge states, going from the
contact $i$, $\mu_i$ is the electrochemical potential of the
contact $i$, $n_i$ is the  number of edge states that are going
from the contact $i$. It is worth to mention here, that edge-state
transport is dissipativeless in the absence of inter-edge
scattering. A finite resistance is arising due to the mixing of
the electrochemical potentials in ohmic contacts.

\subsection{Experiments at low imbalance}
\begin{figure}
\centerline{\includegraphics[width=\columnwidth]{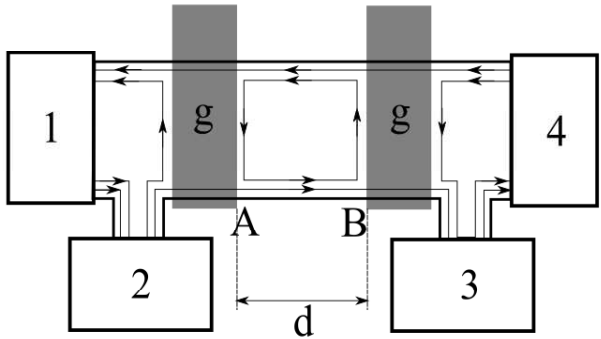}}
\caption{Hall-bar geometry with two crossing gates
Numbered rectangles denote ohmic contacts. Shaded areas are gates
evaporated onto the sample. The structure of edge states is shown for the filling
factors $g=1$ under the gate and $b=2$ in the rest of the sample. (After Ref.~\protect\onlinecite{mueller92})
\label{fig1}}
\end{figure}
Experimental verification of B\"uttiker's formalism was performed
mainly in the Hall-bar geometry with crossing gates (see
Fig.~\ref{fig1}). In this geometry, a sample with two current
leads (1 and 4 in Fig.~\ref{fig1}) and several potential contacts
(2 and 3 in Fig.~\ref{fig1}) was crossed by one or several gates.
Reducing the electron concentration under the gates to the smaller
than the bulk filling factor  results in a nonzero voltage between
potential  contacts in the quantum Hall effect regime. The result
can easily be explained in terms of edge states: in the absence of
gate voltage, two edge states leave contact 2 and the same states
arrive at contact 3, see Fig.~\ref{fig1}. Because no current flows
through the potential contacts, their electrochemical potentials
are equal, leading to zero voltage drop between contacts. If the
filling factor beneath the gates is reduced, then some of the edge
states are reflected at the gate boundary while the others pass
through, which leads to a more complicated set of electrochemical
potentials of the contacts.It can be calculated from Buttiter
formulas (~\ref{butt_form}) for the particular situation.
Moreover, this geometry allows to model and study  various effects
considered by B\"uttiker. A very comprehensive review of experiments
in this geometry is made in Ref~\cite{haug93}.

In the region between the gates in Fig.~\ref{fig1}, one of the
edge states starts from beneath the gate and the other approaches
the gate along the gate edge. Their electrochemical potentials are
different in general. Further along the sample edge, the
electrochemical potentials of these states come to an equilibrium
due to the electron transport between them, i.e., across the
sample edge. Thus, the transport effects between the edge states
can be studied if the mixing of states in the contact can be
excluded, i.e., if a second crossing gate is used as a detector of
the final electrochemical potential of the edge state, as shown in
Fig.~\ref{fig1}. Using the B\"uttiker formalism (\ref{butt_form}),
it is easy to see that the measured resistance is
\begin{equation}
R_{14,23}=\frac{h}{e^2}\left[1+exp\left(-\frac{2d}{l_{eq}}\right) \right]^{-1},
\label{mueller_form}
\end{equation}
where $l_{eq}$ is the phenomenological equilibration length
between the edge states. It can therefore be found from the deviation in the
measured resistance from the quantized value.

Experimental data obtained by various groups (see, e.g.,
Ref.~\cite{mueller92}) have shown that the equilibration length
between spin-split edge states can reach 1 mm at low temperatures
and is of the order of 100~$\mu$m for ones separated by a
cyclotron splitting. This difference is caused by the fact that
the spin flip accompanying the electron transfer is hampered at
the edge of the sample: there are no magnetic impurities in
perfect heterostructures, and the spin flip is due to the
spin-orbital and hyperfine
interactions~\cite{mueller92,khaetskii92,dixon97,komiyama02}.

We note that such experiments provide information about the
equilibration processes only for a small imbalance (a small
difference in electrochemical potentials in comparison to the
spectral gaps) between the edge states. In fact, any initial
imbalance can be applied, but  to have measurable deviations from
the quantized value in (\ref{mueller_form}), $l_{eq} \sim d$
should be fulfilled. Thus, the resulting processes are at low
imbalances, between the edge states that are  practically in equilibrium. The
physical origin is the following: in the Hall-bar geometry we
study transport across the edge as the some correction to the
constant transport along the edge. At high imbalances this
correction is too small to be investigated. Thus, Hall-bar
geometry is very suitable to study undisturbed situation at the
sample edge. At high imbalances, however, edge reconstruction can
be anticipated, following a lot of interesting physical effects.
For the investigations in this regime we should switch to the
Corbino topology.

\subsection{Edge structure in real samples}
\begin{figure}
\centerline{\includegraphics[width=\columnwidth]{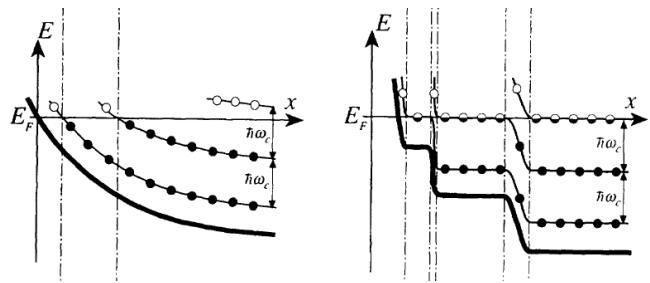}}
\caption{Edge structure for the smooth edge potential  (After
Ref~\protect\onlinecite{shklovsky92}). Left panel: simple one-particle
picture. Right panel: edge reconstruction because interaction
effects. \label{fig2}}
\end{figure}
Before describing the special features of the
transport between the edge states in the case of an arbitrary
imbalance, where the details of the edge structure manifest
themselves, we give a detailed description of the structure of
a real (in most cases, etched or electrostatic) edge of a sample.

The edge potential is \textit{smooth} smooth if it varies on a
length scale much larger than the magnetic length. This is true
for  usual experimental realizations in the integer quantum Hall
effect regime, e.g. etched mesa edge or the electrostatic confinement, because of high values of the spectral gaps and the long-range character of the Coulomb potential.   In the
case of a smooth potential, the bottom of the two- dimensional
subband rising up in approaching the edge of the sample, and the
Landau levels follow the subband bottom (Fig.~\ref{fig2}). At any
point, a local filling factor (the number of filled levels) can be
introduced, which varies from its initial value in the bulk of the
sample to zero at the edge. A change in the local filling factor
occurs each time a Landau level crosses the Fermi level, see
Fig.~\ref{fig2}, left. Chklovsky et. al.~\cite{shklovsky92} took
the electron-electron interaction into account in the mean-field
approximation. It turned out that one-dimensional intersections of
the Fermi and Landau levels are transformed into finite-width
strips (in a certain region, the Landau level is `pinned' to the
Fermi level, see Fig.~\ref{fig2}, right) where the local filling
factor gradually varies, and the edge of the electron system is an
alternating sequence of compressible and incompressible strips of
electronic liquid. The incompressible strip width is determined by
the energy gap between the corresponding Landau levels. The strips
of compressible and incompressible electron liquid can be observed
directly in spatially resolved techniques, see, e.g.,
Ref.~\cite{image02}.

In this case, we should answer the question about the current
distribution over the sample. It was clearly showed by
Thouless~\cite{thouless93} that dissipationless (diamagnetic)
currents flow in regions with a potential gradient because the
group velocity in such areas is nonzero. It means that they are
concentrated in the incompressible strips at each sample edge. If
the electrochemical potentials of the edges are different, the
current in one direction exceeds the opposite current by exactly
the value determined by the difference in the electrochemical
potentials of the edges. This justifies the validity of the
B\"uttiker formalism, which is sensitive only to integral
characteristics, such as the electrochemical potentials of the
edges and the matrix of scattering coefficients `from contact to
contact.' This consideration pertains to the current along the
edge of the sample. The current running across the edge and
equilibration of the edge states is determined by tunnelling
through incompressible strips and diffusion in compressible ones.

Taking these considerations into account, we can reformulate the
definition of an edge state as a \textit{compressible} strip.
This provides a clear definition for the electrochemical potential
of an edge state and keeps our consideration consistent with the
above considerations, where an edge state was defined as the
intersection of a Landau level with the Fermi level.

\section{Transport between edge states at high imbalance in the integer quantum Hall effect
regime}\label{IQHE}
Most probably, a quasi-Corbino geometry in
combination with the technique of a crossing gate was first
proposed in Ref.~\cite{weiss92}. But the first experimental
results appeared only ten years later~\cite{alida02}, when a
measurement method appropriate for obtaining interpretable results
was developed and the experimental difficulties arising in such
measurements were overcome. By that time, the idea of applying the
Corbino geometry had been thoroughly forgotten, and the authors of
Ref.~\cite{alida02} had to develop the sample geometry anew.

\subsection{Quasi-Corbino experimental geometry}
\begin{figure}
\centerline{\includegraphics[width=\columnwidth]{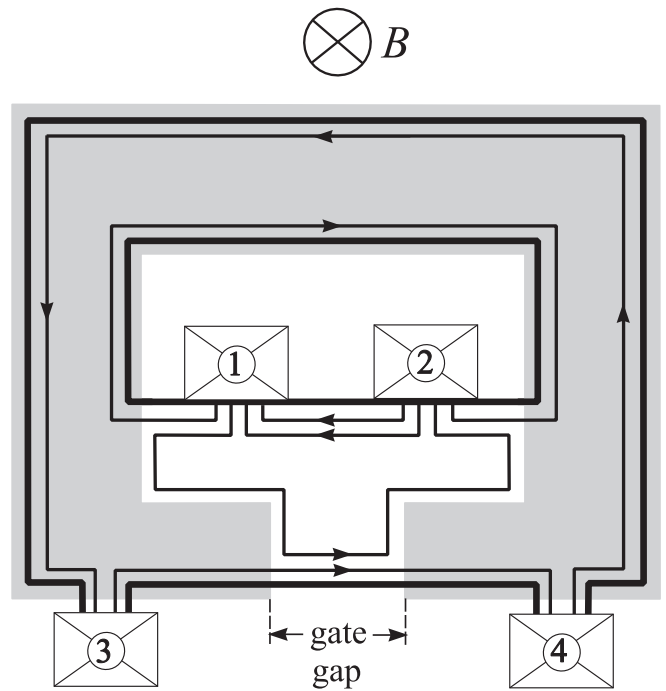}}
\caption{ Sketch of a sample in the quasi-Corbino geometry (After
Ref.~\onlinecite{relax04}). Bold lines show the inner and outer edges of
the mesa, crossed rectangles with numbers denote ohmic contacts,
the shaded area is the gate. The structure of the edge states is
shown in the case where the filling factor is $g=1$ under the gate
and $\nu=2$ outside the gate. \label{fig3}}
\end{figure}
In the quasi-Corbino
geometry (see Fig.~\ref{fig3}), the sample has the rectangular shape
with an etched region at the center, which creates two
independent edges not connected topologically. Ohmic contacts are made to the
two-dimensional electron gas (2DEG) at both edges
A metal gate is evaporated on the top of the sample. It surrounds the internal etched area, leaving only a
T-shaped region of the two-dimensional gas between the
outer and inner boundaries uncovered (see Fig.~\ref{fig3}).

By partially depleting the 2DEG under the gate, different filling
factors in the T-shaped region ($\nu$) and under the gate ($g$)
can be achieved, $g<\nu$. Some part of the edge states are
reflected at the boundary of the gate and go along the gate to the
other boundary of the sample. At the inner boundary of the sample
not covered by the gate (the `bar' of the T), all edge channels
are in equilibrium due to the macroscopic size and several ohmic
contacts. At the outer boundary of the sample, the area not
covered by the gate (the `leg' of the T), is of several
micrometers in size. It is much less than the equilibration length
at low temperatures, and there are no ohmic contacts here. Thus,
if a voltage is applied to a pair of contacts at the inner and
outer edges, a difference in electrochemical potentials appears
between the edge states at the outer edge of the sample, in the
area not covered by the gate, i.e. in the gate-gap region.

To obtain $I-V$ characteristics of the transport between  edge
states in the gate-gap,  4-point configuration is used. A
\textit{dc} current is applied  between a pair of inner and outer
contacts and  the resulting \textit{dc} voltage  is measured
between another pair of inner and outer contacts.
 (Due to the existence of a preferred direction
determined by the magnetic field, there are four principally
different combinations of contacts.) Four-point configuration
allows to eliminate contact effects. The obtained results were
qualitatively confirmed by direct measurements of two-terminal
$I-V$ characteristics.

This geometry offers many degrees of freedom to a researcher. By
varying the filling factor in the gate-gap area with the help of a
magnetic field, the total number of interacting edge channels can
be changed; by varying the filling  factor beneath the gate with
the help of the gate voltage, the channels can be divided into
groups to which the difference in electrochemical potentials is
applied and between which the current flows. In particular, the
transport between edge states separated by a spin or cyclotron gap
or, in double-layer structures, by symmetric-antisymmetric
(isospin) splitting can be studied in the IQHE regime.

\subsection{Switching from the low imbalance to the high imbalance case}
\begin{figure}
\centerline{\includegraphics[width=\columnwidth]{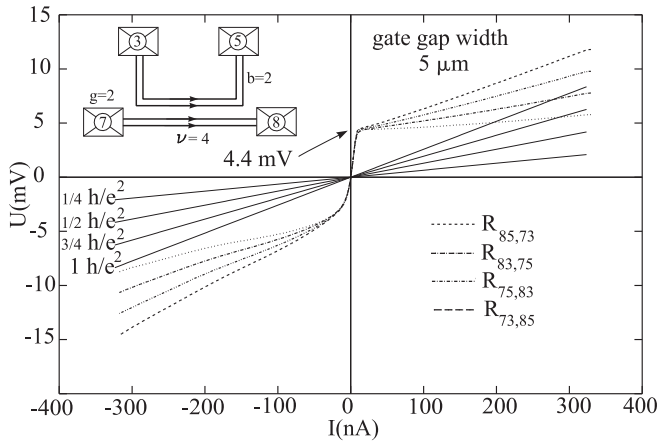}}
\caption{Current-voltage characteristics at a high temperature
(straight lines, 4K, complete equilibration) and a low temperature (non-linear curves, 30 mK,
non-equilibrium regime) for the transport between cyclotron-split
edge states (After Ref.~\protect\onlinecite{alida02}). \label{fig4}}
\end{figure}
Transformation of the $I-V$ curves was demonstrated with
decreasing the temperature~\protect\cite{alida02}. At a high
temperature (4 K), the $I-V$s are linear, with the slope exactly
corresponding to its equilibrium value obtained from the B\"uttiker
calculation (\ref{butt_form}) for various combinations of ohmic
contacts (see Fig.~\ref{fig4}). This fact can be explained by the
small value of the equilibration length (compared to the size of
the interaction area length, i.e. gate-gap width) at this
temperature. As the temperature decreases to 30 mK, the
equilibration length grows dramatically~\cite{mueller92}, and the
system enters the regime of strong imbalance. The current-voltage
characteristic becomes strongly nonlinear and asymmetric (see
Fig.~\ref{fig4}), with a pronounced threshold behavior of the
right-hand branch (corresponding to positive currents when the
inner contact is grounded). Above the threshold, this branch is
linear, while the left-hand branch has no threshold and remains
nonlinear. A similar transformation of the $I-V$ was observed at
low temperatures~\cite{fraceq06} as a result of \textit{in situ}
varying the length of the interaction area.

\subsection{Interpretation of the non-linear current-voltage characteristics}
\begin{figure}
\centerline{\includegraphics[width=\columnwidth]{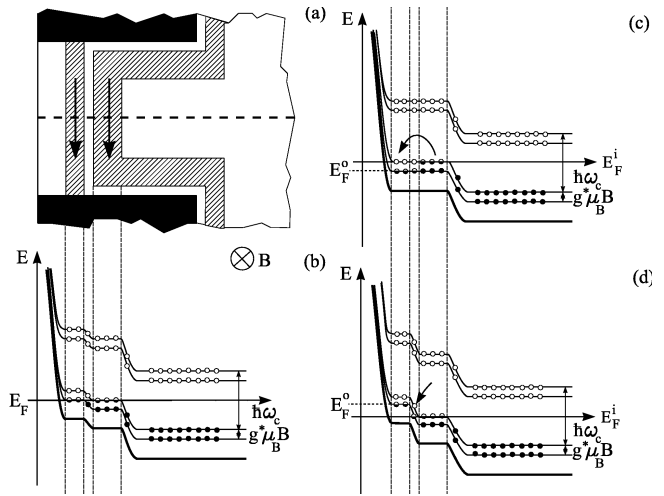}}
\caption{  (Left panel) The structure of the sample edge in the
interaction area of edge states in equilibrium. The bulk filling
factor $\nu$ is equal to 2. Two spin- split energy levels reach
the edge and form a structure of compressible and incompressible
strips. (Right panel) The structure of the sample edge in the
interaction area of edge states in the case of a voltage applied
between compressible strips. At positive voltage, because the
electron charge is negative, the potential barrier between the
edge states reduces down to the  flat-band situation (c). At
negative voltage, the barrier grows and deforms (d). (After
Ref.~\protect\onlinecite{alida02}). \label{fig5}}
\end{figure}

The above mentioned non-linear $I-V$ curves can only be
explained~\cite{alida02} by means of the smooth edge model, in
which the edge is represented by alternating strips of
compressible and incompressible electron liquids (see
Fig.~\ref{fig2} and Fig.~\ref{fig5}, which -for simplicity-
discusses the case $\nu=2)$).  At bulk filling factor $\nu=2$, the
interaction area near the outer boundary contains two compressible
strips separated by the incompressible strip with the local
filling factor $g=1$. The electrochemical potential of each
compressible strip is determined by the electrochemical potential
of the corresponding (inner or outer) ohmic contact. If a voltage
is applied to a pair of contacts situated at different edges, a
difference in electrochemical potentials drops within the
incompressible strip between the two compressible ones and affects
the distribution of edge potential in it. For instance, at
positive voltages (inner contacts grounded), the potential barrier
between the edge states decreases and completely disappears when
the voltage is equal to the corresponding spectral gap (see
Fig.~\ref{fig5} (c)). This leads to a dramatic growth in the
current at this voltage and to a complete equilibration between
the edge states at larger potential differences. At negative bias
voltage, the potential barrier increases, which leads to the
appearance of a strongly nonlinear $I-V$ branch.

\subsection{Spectral investigations}
Thus,  the energy gap between the edge states can be found from
the position of the threshold voltage on the right-hand (positive)
$I-V$ branch. It turns out that the gap is equal to the bulk value
of splitting between the corresponding energy levels. This fact
was first demonstrated for cyclotron gaps~\cite{alida02}, which
justifies using the smooth-edge model and experimentally confirms
the smoothness of an etched edge in the IQHE regime. (All the
arguments for the $I-V$ should also be valid in the case of a
sharp edge, but the measured gap is then much larger than the bulk
value of the splitting). For sufficiently pure samples, it was
shown in~\cite{fracbutt} that the gap between spin-split edge
states corresponds to the bulk exchange-increased Lande
factor~\cite{dolgopolov97}.

\subsection{Equilibration at the edge}
In addition to spectroscopic studies, the process of equilibration
was  studied in Ref.~\cite{alida04}, with the initial values of
imbalance exceeding the spectral gap in the transport between
cyclotron-split edge states. In this experiment, the slope of the
linear (above- threshold) part of the $I-V$ right-hand branch was
studied (see Fig.~\ref{fig4}). It turned out that for strongly
nonequilibrium edge states, not the whole difference of
electrochemical potentials but only the part exceeding the
spectral gap can be redistributed.

Furthermore, the B\"uttiker formalism~\cite{buttiker88} was modified
 by explicitly
introducing a local characteristic of the transport between the
edge states instead of the integral matrix $T_{ij}$. Namely, the
local transport parameter $\alpha$ was defined as the ratio of the
distributed difference in electrochemical potentials between the
edge states and the difference in electrochemical potentials
allowed for redistribution. This single parameter a is universal:
it fully describes the slopes of linear parts of the $I-V$ for any
combination of the contacts and depends only on the physics of the
transport between edge states. Numerical values of $\alpha$
indicate the extent to which equilibrium is established between
the edge states.

\subsection{Spin-flip transport: creation of dynamic nuclear polarization}
\begin{figure}
\centerline{\includegraphics[width=\columnwidth]{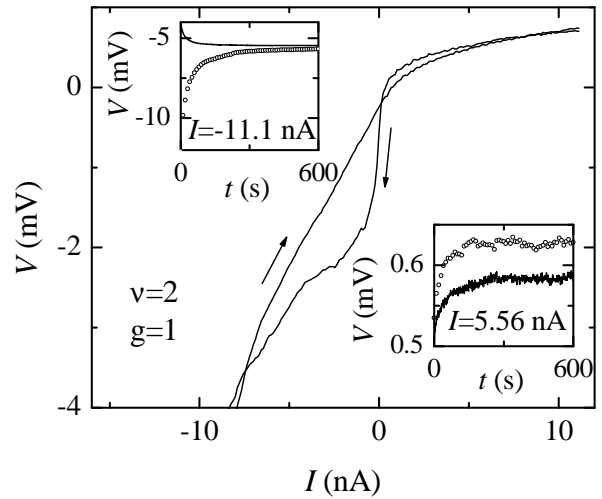}}
\caption{Hysteresis for the $I-V$ for the spin-
flip transport and relaxation curves (After Ref.~\protect\onlinecite{relax04}).
\label{fig6}}
\end{figure}
The transport between spin-split edge states should be accompanied
by the electron spin flip. The spin flip is mainly provided by the
spin-orbital interaction~\cite{mueller92}, but part of the
electrons participate in the so-called flip-flop process: due to
the hyperfine interaction, the spins of the electron and the
nucleus are flipped simultaneously. This process, even at
relatively high temperatures, leads to the creation of an area
with a dynamic polarization of nuclear spins, in which the static
polarization by the external magnetic field is
inessential~\cite{dixon97,komiyama02}.

Creation of dynamic nuclear polarization has been studied in the
strongly nonequilibrium case in Ref.~\cite{relax04}, where the
$I-V$s were measured for the transport between spin-split edge
states in the quasi-Corbino geometry. Under these conditions, the
current-voltage characteristics exhibit a considerable hysteresis,
especially pronounced in the left- hand (negative) branch (see
Fig.~\ref{fig6}). Comparison with the $I-V$ obtained for transport
without the spin flip (through a cyclotron gap) in the same field
and with the same degree of disorder showed that the hysteresis is
not related to the spurious transient effects such as recharging
of the sample from the contacts. It was shown in
Ref.~\cite{relax04} that the hysteresis is caused just by the
dynamic polarization of the nuclei in the interaction area of the
sample. Indeed, the effective Overhauser field arising in this
case influences the spin splitting, which determines the potential
barrier between the edge states. This affects on the current for
all electrons, and not only for those whose spin flipping is
caused by the flip-flop process; as a result, a noticeable
hysteresis of the $I-V$ occurs.

In addition, relaxation processes investigated in Ref.~\cite{relax04}
 revealed two typical relaxation times,
of the order of 25 and 200~s (see the insets in Fig.~\ref{fig6}).
The first time corresponds to the creation of the dynamic nuclear
polarization area at a certain stage of the transport between the
edge states, and the second relates to the development of a stable
area where nuclear spins in the sample are polarized due to the
competition between the nuclear spin diffusion and the escape of
the spins from the system.

It was also demonstrated~\cite{axel05} that the flip-flop
mechanism can be  reversed. After establishing a local dynamic
nuclear polarization region, the externally applied current is
switched off, and the sample exhibits an output voltage, which
decays with a time constant typical for the nuclear spin
relaxation.

\subsection{Edge states in the double quantum wells. Topological defects in the edge state structure}

More complicated for investigation are tunnel-coupled double
electron layers, or double-layer systems, which are usually
realized in double quantum wells separated by a tunnel-transparent
barrier.Because of the tunnelling between the layers, the bulk
spectra of such systems are already rather
complicated~\cite{davies96,dqw99,tilt00,japan99,pellegrini97,vadik00}.
At the edge of the sample, the Fermi level becomes the same for
edge states that originate, in the general, from different parts
of the quantum well or from subbands, depending on the quantum
well symmetry.

In the symmetric case, in addition to the cyclotron and spin
splitting, a symmetric-antisymmetric splitting appears, which is
smaller than the Zeeman splitting in strong fields and exceeds it
in weak fields~\cite{japan99,pellegrini97,vadik00}.. In
intermediate fields, where these splitting values should be
comparable, a new phase, the so-called antiferromagnetic one,
appears due to the electron-electron
interaction~\cite{DasSarma97}.. A bulk transition into this phase
from the range of weak fields was observed in Ref.~\cite{vadik00}
and from the range of strong fields in Ref.~\cite{pellegrini97}.
Thus, singularities can be expected in the transport between edge
states in the vicinity of the bulk phase transition point. Such
transport singularities were observed in Ref.~\cite{dqwedge04},
where the incompressible strip separating edge states was
demonstrated to disappear near the bulk phase transition point.

An important fact, established in Ref.~\cite{dqwedge04}, is that the
structure of edge states always corresponds to the structure of
the bulk spectrum and follows even its complicated
transformations. It was investigated by mapping the energy gaps at the edge using $I-V$ spectroscopy, while the system approach the phase transition point.

\begin{figure}
\centerline{\includegraphics[width=\columnwidth]{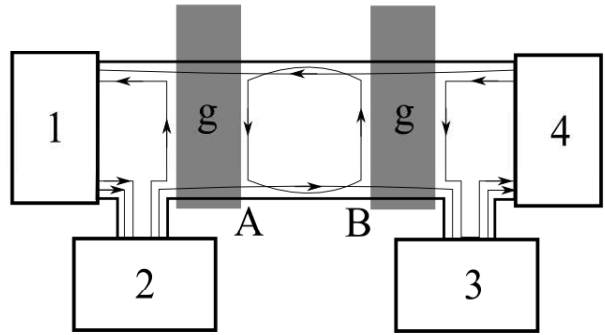}}
\caption{The simplest example of the topological defects in the edge states structure
(After Ref.~\protect\onlinecite{bauer94}).
\label{fig7}}
\end{figure}

The Pauli principle does not forbid the intersection of edge
states corresponding to different quantum numbers (see,  for
example, Fig.~\ref{fig7}). Such intersections were called defects
in the topological structure of edge states, or topological
defects. The possibility of the existence of such defects was
shown theoretically in Refs~\cite{demsley94,bauer94}.

The only way to detect topological defects is by studying the
transport between edge states, which can be most conveniently done
in the quasi-Corbino geometry~\cite{alida02}. The existence of a
gate in this geometry, in particular, allows changing the symmetry
of the quantum well, and hence the energy spectrum under the
gate~\cite{dqw99,tilt00}. Since the structure of edge states
corresponds to the structure of the bulk
spectrum~\cite{dqwedge04}, the quasi-Corbino geometry allows to
realize topological defects in the structure of edge
states~\cite{topdefect05}: (i)If the well is asymmetric in the
interaction area of the edge states, then edge-state electrons are
fully described by the spin and isospin (layer number)
orientations~\cite{dqw99}. Electrons injected from beneath the
gate can also come either from an isospin-polarized state or from
a mixed one. In the latter case, because the injection occurs with
isospin conservation, the electrons are distributed among the edge
states. This results in the intersections of edge states and in
the equilibration of electrochemical potentials for all edge
states in the interaction area, which manifests itself in the
perfect linearity of the $I-V$ in a normal magnetic field. (ii) If
a tangential field is applied, the states in the interaction area
become isospin- mixed~\cite{tilt00}, and the topological defects
disappear. This leads to a strongly nonlinear $I-V$, usual for the
inter-edge-state transport. In this way, the existence of
topological defects and the possibility of controlling their
creation and disappearance were demonstrated in
Ref.~\cite{topdefect05}.

\section{Transport between edge states at high imbalance in the fractional quantum Hall effect regime}

\subsection{Laughlin's wavefunction and the composite fermion hypothesis. Edge states  in the FQHE}
In the fractional quantum Hall effect regime, the system has to be
treated as  a large number of strongly interacting particles, and
therefore no method exists for exactly solving the problem with
the real Hamiltonian. The interaction results in a rearrangement
of the ground state of the system of particles, and the new ground
state cannot be obtained from the perturbation theory as a small
correction to the interaction- free state.  Two approaches turned
out to be efficient for the description of such a liquid: the
method  of a trial ground-state wave function (the Laughlin
approach~\cite{laughlin83}) and the mean-field
method~\cite{jain89} (based on the hypothesis of composite
fermions).

According to MacDonald~\cite{macdonald90} there are collective
gapless excitation modes at  the sample edge in the FQHE regime,
that he defined as edge states. The structure of the excitation
spectrum was shown to correspond to the  structure of the
Laughlin's ground state at the particular filling factor.  For
example, the state with $\nu=2/3$, according to Laughlin, is
constructed as a quasi-hole state on the background of a
completely filled lowest Landau level. Correspondingly, in this
case, edge spectrum consists from the branches, originating from
quasi-holes and electrons correspondingly. He attributed a current
to each excitation branch $I=(e^*/h)\Delta\mu$, where $e^*=e\nu$
is the effective charge of the branch (1/3 and -1 in the above
example) and completed the construction of a FQHE B\"uttiker
formalism~\cite{buttiker88} by introducing the transmission matrix
$T_{ij}$, \cite{macdonald90}.

The so-defined edge excitations are one-dimensional and in the
FQHE regime the inter-electron interaction must be consistently
taken into account. This was done in the theoretical works by
Wen~\cite{wen91}, who applied the Luttinger
model~\cite{luttinger63}  of a one-dimensional interacting liquid
to this problem and demonstrated that collective excitations with
a gapless spectrum do exist at the edge and their structure is
indeed determined by the hierarchical structure of the bulk ground
state. Physically, these excitation branches correspond to
different modes of edge magnetoplasmons (see,
e.g.,~\cite{aleiner94} that we will need below). We note that an
edge state in the FQHE regime is probably the only exact
realization of a chiral Luttinger liquid model: the edge creates
the one-dimensionality of the system, the bulk states form an
infinite reservoir, which is necessary in the Luttinger model, and
the magnetic field determines a preferred direction providing the
chirality of the electron liquid. Therefore, the investigation of
collective excitations in the FQHE regime allows studying a rare
example of a non-Fermi electron liquid.

Since the transport along the edge is  determined by the bulk
filling factor and edge electrochemical
potentials~\cite{macdonald90}, the only way to study Luttinger
liquid effects is the transport across the edge. Wen~\cite{wen91}
has shown theoretically that the tunnel density of states has
power-law behavior in the FQHE regime, $D(E)\sim E^{1/g-1}$ with
$g=1/\nu$ for the filling factors  $\nu$ from the principal
Laughlin sequence. It was also shown in~\cite{kane92} that there
are universal scaling relations for the temperature dependence of
the tunnel density of states $D \sim T^{1/g}$.   These results
have also been confirmed in the approach of composite
fermions~\cite{shytov98}.

    In the experimental study of tunnelling into the edge, it
must be ensured that the $I-V$ nonlinearity is caused precisely by
the excitation of collective modes and not by the deformation of
the edge potential. For this, the so-called cleaved edge
overgrowth technique~\cite{chang96} is used. Experiments in
Refs~\cite{chang96,grayson98} demonstrated power-law $I-V$s in the
case of tunnelling into the edge, as well as temperature scaling
of these diagrams with the exponents close to the predicted
ones~\cite{wen91,kane92} for the filling factor $\nu=1/3$. The
experiment and the theory give considerably different
results~\cite{grayson98} outside the vicinity of $\nu=1/3$, which
might be caused by a structure of compressible and
incompressible strips forming at the edge~\cite{chang96}.

Numerical calculations based on the Laughlin
wave function~\cite{chamon94} and in the framework of the composite-
fermion approach~\cite{chklovskyCF95} showed that the structure of strips of
an incompressible and compressible electron liquid already
appears at the edge width as small as five or six magnetic
lengths. In other words, all real potentials (such as, for
instance, the most common potential of a mesa etched edge)
satisfy this condition.  The situation is still not so clear for the  cleaved edge overgrowth (CEO) samples~\cite{chang96,grayson98}, which are the best candidates for the sharp edge realization.
On the one hand, there are signs of the compressible-incompressible strips formation in high magnetic fields~\cite{chang96}, supporting the above mentioned calculations~\cite{chamon94,chklovskyCF95} . On the other hand, there are sings of the sharp edge situation~\cite{huber05} in low magnetic fields, at much higher magnetic length. This difference could also occur from the progress in the CEO samples preparation over a decade. In this Review we will concern only the high-field limit, as the most appropriate for the FQHE regime.

For a smooth potential, the bottom of the two- dimensional subband
increases in the vicinity of the edge and the electron
concentration decreases. Hence, a local filling factor can be
introduced, which varies from the bulk value to zero in
approaching the edge of the sample. Beenakker~\cite{beenakker90}
showed that for a sufficiently pure system and the FQHE existing
with such local filling factors, finite- width incompressible
strips corresponding to these filling factors appear on the edge.
In the FQHE regime, therefore, similarly to the IQHE case, a
smooth edge consists of alternating strips of compressible and
incompressible electron liquid. The difference from the integer
case lies in the fact that it is now impossible to introduce a
system of Landau levels bent at the edge, because everything
occurs on the last (single) Landau level. It can only be asserted
that there is no gap in the compressible strips, while a gap
corresponding to the electrochemical potential between the ground
and the excited states occurs in the incompressible strips. This
gap shrinks at the edges of each incompressible strip.
Dissipation-free current, similarly to the IQHE case, is carried
by the ground state and, because the `excess' current is
concentrated near the edge of the incompressible area in the
absence of equilibrium, it can be described as an edge current.

As in the integer case, the analogue of the B\"uttiker
formalism can now be introduced as~\cite{beenakker90}:
\begin{equation}
I_i=\frac{e}{h} \left(\nu_i\mu_i+\sum_{j\ne i} T_{ij}\mu_j\right)
\label{butt_frac},
\end{equation}
where $I_i$ is the current carried by the edge states coming out
of contact $i$, $\mu_i$ is the electrochemical potential of
contact $i$, and $\nu_i$ is the maximum filling factor for
incompressible strips coming from contact $i$. It is easy to see
that equation~\ref{butt_frac} contains B\"uttiker
formula~\ref{butt_form} as a special case of integer $\nu_i$, as
well as MacDonald's result~\cite{macdonald90} for a sharp edge
potential, because $e^*=e\nu$. This indicates that B\"uttiker
formalism is a rather general integral relation, which is
independent of the details of the edge structure. Similarly to the
integer case, to check the formalism one should place a crossing
gate on the sample. Such experiments showed a perfect agreement
between the calculation and the measurement~\cite{crossgates85}.

While strips of incompressible electron liquid exist at a smooth
edge of a two-dimensional electron system in the FQHE regime,
collective modes appear near the boundaries of these
strips~\cite{chamon94}. In addition, because the edges of the
strips are close (both to each other and to the edges of
neighboring strips if the potential is not very smooth), and the
electric fields are long-range ones, these modes
interact~\cite{vignale96}. Therefore, collective excitations on a
smooth edge in the FQHE are most similar to neutral magnetoplasmon
modes, which were first proposed for the IQHE
regime~\cite{aleiner94}. As a result, in the case of tunnelling
into a smooth FQHE edge, the exponent of the tunnel density of
states and hence the $I-V$s become dependent on the real shape of
the edge potential~\cite{vignale96}, although the $I-V$ maintains
its power-law behavior, which was demonstrated in
Ref.~\cite{grayson98}.

\subsection{Transport across the incompressible strip at high imbalance}
As it was shown before, there are two major problems in
investigation of  collective effects at the smooth sample edge in
FQHE regime: (i) a presence of the structure of compressible and
incompressible strips, which entangle different collective modes
at the strip edges; (ii) deformation of the edge potential by the
applied voltage $V$ while measuring $I-V$ curves. This affects the
 $T_0(V)$ dependence, where $T_0$ is one-particle barrier
transmittance, and, therefore, makes it difficult to separate
one-particle effects and the collective ones. The former problem
can be removed by separate contacting to compressible strips
across a single incompressible one. The latter problem demands the
high-imbalance regime. Indeed, collective effects can be selected
in two limiting cases. The first is where the bias potential is so
small compared to the potential barrier that it does not deform it
(This regime was realized in Ref.~\cite{grayson98}, but without
separate contacting to the strips). The second case is where the
bias potential is  large in comparison with the barrier width. The
barrier is deformed and other deformation has no effect on
transport across it. (This fact can be easily understood in a
triangular barrier approximation). We note that the second case is
easier to realize from the experimental standpoint, and it can be
better controlled. Thus, experiments in the quasi-Corbino geometry
are needed to study collective effects at the smooth sample edge.

\begin{figure}
\centerline{\includegraphics[width=\columnwidth]{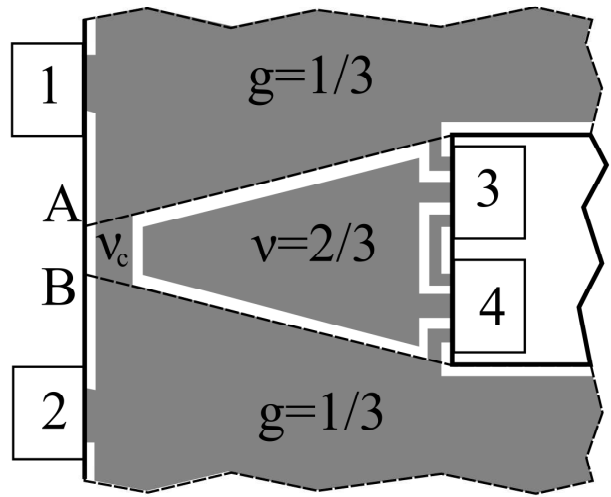}}
\caption{Schematic diagram of  the active region of the sample in
the quasi-Corbino geometry. The etched mesa edges are shown by
solid lines, the dashed lines represent the split-gate edges. The
gate-gap region at the  outer mesa edge is denoted as AB. Light
gray areas are the incompressible regions at filling factors $\nu$
(in the bulk), $g$ under the gate ($g<\nu$), and at local filling
factor $\nu_c$ ($\nu_c=g$) in the incompressible stripe at the
mesa edges. Compressible regions (white) are at the
electrochemical potentials of the corresponding ohmic contacts,
denoted by bars with numbers. (After
Ref.~\protect\onlinecite{fracdens06}) \label{fig8}}
\end{figure}

Figure~\ref{fig8}  shows the structure of compressible and
incompressible electron liquid strips near the gate gap of the
sample in the quasi-Corbino geometry for the simples situation of
filling factors $g=1/3$ under the gate and $\nu=2/3$ outside it.
This scheme is based on the data of magnetoresistance and
magnetocapacitance measurements. For instance,  measuring the
magnetoresistance in the quantum Hall effect regime allows finding
the field corresponding to the filling factor $\nu=2/3$ in the
part of the sample not covered by the gate. Further, the
capacitance between the two-dimensional system and the gate should
be measured while decreasing  the electron concentration under the
gate. This allows finding the FQHE fractional filling factors
manifested in the given sample at given magnetic fields due to a
decrease in the electron concentration. Because approaching the
edge is also accompanied by a decrease in the electron
concentration for the smooth edge potential, we can be sure that
incompressible strips appear at the edge of the sample at the same
filling factors that were observed when decreasing the electron
density beneath the gate. For example, in the sample described in
Ref.~\cite{fracbutt}, at the bulk filling factor $\nu=1$,
incompressible strips appear in the vicinity of the edge at local
filling factors $2/3$ and $1/3$. Choosing the filling factor under
the gate to coincide with one of these values, we choose the
incompressible strip for which the transport is studied.

    Similarly to the IQHE case, obtaining current-voltage
characteristics is the basic tool in the study of the transport.
Measurement of the transport through an incompressible
strip can be carried out in two ways: by fixing the current or
by fixing the voltage. Because the FQHE is especially sensitive
to the quality of the samples and the ohmic
contacts, to check the
results for reliability it is necessary to see whether the data of
both above-described methods of the $I-V$ measurement
agree in each particular case. In addition, it is necessary to
independently estimate the resistance and the quality of the ohmic
contacts using magnetoresistance measurements, to use
various samples and different methods to cool them, and to
compare the results with the ones known for the IQHE
regime.

\begin{figure}
\centerline{\includegraphics[width=\columnwidth]{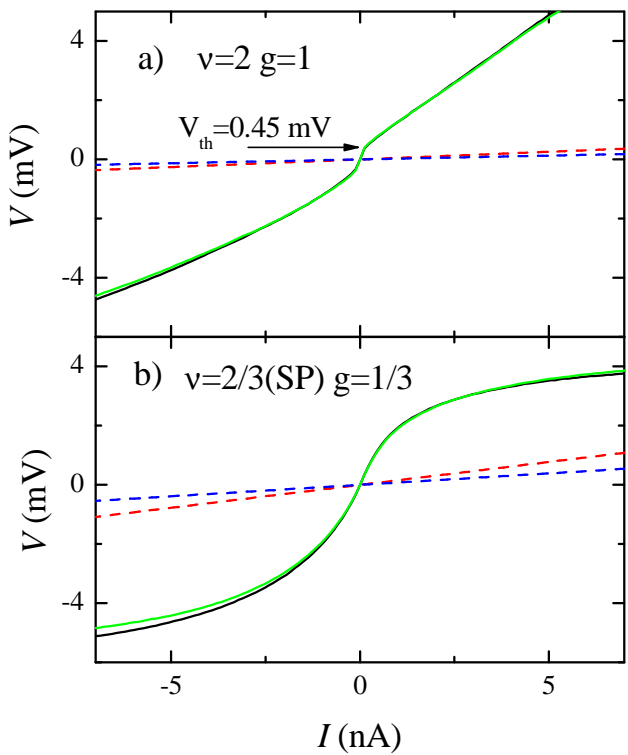}}
\caption{ $I-V$ curves for integer filling factors $\nu=2, g=1$
(a) and fractional ones $\nu=2/3, g=1/3$ (b) for two different
contact configurations for for a sample with an extremely narrow
gate-gap width $L_{AB}=0.5 \mu$m. Equilibrium lines (with
$R_{eq}=2;1 h/e^2$ (top) and $6;3 h/e^2$ (bottom)) are  shown.
Magnetic field $B$ equals to 1.67~T for integer fillings and to
5.18~T for fractional ones. (After
Ref.~\protect\onlinecite{fracdens06}). \label{fig9}}
\end{figure}
     We consider a current carried across an incompressible
strip, shown in Fig.~\ref{fig8}, depending on the equilibration
length $l_{eq}$:
\begin{equation}
I=R_{eq}^{-1} V (1-exp(-L_{AB}/l_{eq})),\label{eq_fracdens}
\end{equation}
For the gate-gap width $L_{AB} \ll l_{eq}$ , the shape of the
current-voltage characteristic directly reflects the behavior of
the equilibration length as the imbalance between the edge states
is varied. In turn, the equilibration length reflects the behavior
of the transition probability $w$ between the edge states, $l_{eq}
\sim w^{-1}$ . The transition probability $w$ can be written as
the single-particle transmittance $T_0$ of the potential barrier
times the tunnel density of states $D$: $w \sim T_0(V) D(V,T)$. As
mentioned before, the barrier can be considered triangular in the
strongly nonequilibrium case, and therefore the single-particle
transmittance, which can be written as $exp(-C\Delta^{3/2}/V)$,
tends to unity when the imbalance exceeds the fractional gap
$\Delta \ll V$. Because the equilibration length for fractional
filling factors and small imbalances are
known~\cite{komiyama02,kouwen90,cunning92} to be of the order of
$10 \mu$m, the samples used in Ref.~\cite{fracdens06} had the
width of their working area $L_{AB}=0.5 \mu$m.

Figure~\ref{fig9} shows examples of $I-V$s for integer and
fractional filling factors in samples with small interaction area.
The differences in the $I-V$s for fractional filling factors from
the well-known $I-V$s for integer filling factors (see above) are
(i) the absence of a threshold; (ii) strong nonlinearity within
the whole voltage range; and (iii) almost perfect symmetry. This
behavior of the $I-V$s is observed for almost all fractional
filling factors.

\begin{figure}
\centerline{\includegraphics[width=\columnwidth]{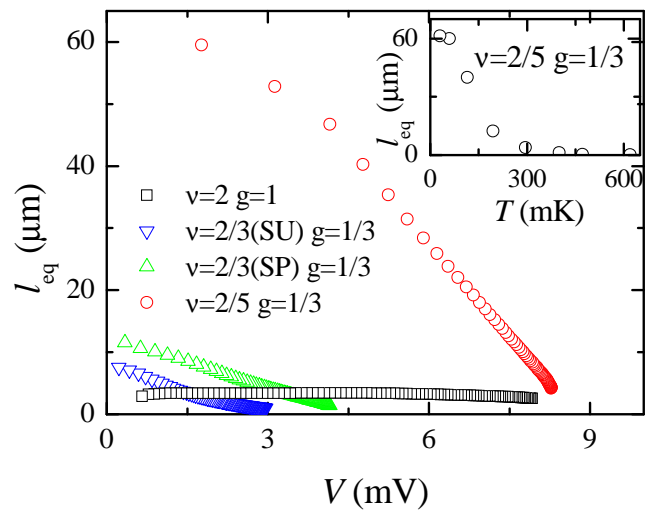}}
\caption{ Equilibration length $l_{eq}$ for different filling
factor combinations as function of the voltage  imbalance $V$
across the incompressible stripe  with corresponding local filling
factor $\nu_c=g$ (see caption to Fig.~\protect\ref{fig8}). Inset
shows an example of the temperature dependence of the $l_{eq}$ for
$\nu=2/5, g=1/3$ ($B=7.69$~T). (After
Ref.~\protect\onlinecite{fracdens06}). \label{fig10}}
\end{figure}
Equilibration lengths calculated by means of
equation~\ref{eq_fracdens} for  various filling factors are shown
in Fig.~\ref{fig10}. For integer filling factors, the $l_{eq}$
behavior corresponds  to the one discussed in section~\ref{IQHE},
known one caused by the deformation of the potential barrier
between edge states, while for fractional filling factors, the
behavior of $l_{eq}$ was studied in Ref.~\cite{fracdens06} for the
first time.

\begin{figure}
\centerline{\includegraphics[width=\columnwidth]{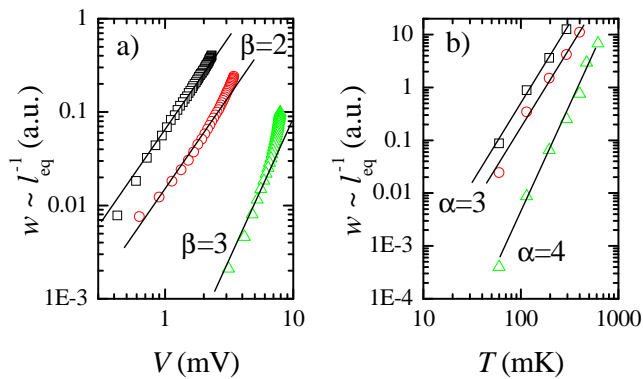}}
\caption{ Transition probability $w \sim l_{eq}^{-1}$ is shown as
a function of the temperature  (a) and  of the voltage imbalance
at $T=30$~mK (b) in logarithmic scales,  demonstrating the
power-law dependencies. The filling factors are $\nu=2/3(SU),
g=1/3$ (squares); $\nu=2/3(SP),g=1/3$ (circles); $\nu=2/5,g=1/3$
(triangles). (After Ref.~\protect\onlinecite{fracdens06}).
\label{fig11}}
\end{figure}
Under the conditions of the experiment~\cite{fracdens06}, the
dependence of the transition probability $w \sim l_{eq}^{-1}$ (see
Fig.~\ref{fig11}) on bias and temperature reflects the dependence
of the tunnel density of states $D$ on these parameters. The
power-law behavior of the transition probability was demonstrated
in Ref.~\cite{fracdens06} (see Fig.~\ref{fig11}); the exponents
found in experiment for the voltage and temperature dependencies
differ by unity, as indeed should be the case under the excitation
of collective modes~\cite{wen91,kane92}. Thus, neutral excitation
modes~\cite{chamon94} do exist at the edges of the incompressible
strip and determine transport across it at high imbalances in the
FQHE regime.

The exponents were found in Ref.~\cite{fracdens06} for the first
time and need a theoretical explanation. They are different for
the filling  factors $\nu=2/3, g=1/3$ and $\nu=2/5, g=1/3$ which
is caused by the collective-mode excitation at the boundary of the
bulk filling factor $\nu=2/5$.The edge of the $\nu=2/5$ bulk
incompressible state is extremely close to the incompressible
strip with local filling factor $\nu_c=1/3$ in this case, because
of $\nu-\nu_c<<\nu$. Thus, we can expect some influence in $D$
also from the edge excitations of $\nu=2/5$ bulk incompressible
state, affecting the exponents in power-low $D(V,T)$. Thus,  the
structure of the collective excitations is more complicated at
$\nu=2/5$,  resembling the acoustic modes predicted in
Ref.~\cite{aleiner94}.

\subsection{Equilibration at the edge and the structure of the excitation spectrum}

Direct measurements of the structure of edge collective
excitations (in the cases where the structure is assumed to be
complicated, see Ref~\cite{macdonald90}) are not realistic: they
would require an independent study of several simultaneously
propagating magnetoplasmon modes~\cite{aleiner94,chamon94}. But an
indirect measurement method is possible. During the equilibration
from an initially strongly nonequilibrium case, the transport
across the incompressible strip involves excitation  of collective
modes~\cite{fracdens06}, which, in turn, establish the edge
potential and therefore influence the
equilibration~\cite{zhulike04}. Such effects are not taken into
account by the single-particle B\"uttiker-Beenakker
theory~\cite{buttiker88,beenakker90}. Hence, the comparison
between the experimental equilibrium resistance and the one
calculated according to B\"uttiker's formalism
(\ref{butt_form},\ref{butt_frac}), can indicate a structure of the
collective excitations.

In Ref.~\cite{fracbutt} equilibration was studied through
incompressible strips corresponding to the local filling factors
$\nu_c=2/3$ and 1/3, at the bulk filling factor $\nu=1$. This
allows investigating equilibration at the same strip structure in
the gate-gap by realizing contacts between different compressible
strips. B\"uttiker's formulas (\ref{butt_form},\ref{butt_frac})
yields the same equilibrium values of resistance for both
combinations of filling factors. However, the
experiment~\cite{fracbutt} showed the equilibrium resistance
values to be different: for the transport through the
incompressible strip with the local filling factor $\nu_c=2/3$,
the slope of the equilibrium curve turned out to be much smaller
than the expected one, while for the transport through the strip
with the filling factor $\nu_c=1/3$, the measured slope was close
to the expected one. In terms of B\"uttiker's formalism,
\textit{smaller} equilibrium slope corresponds to an excess charge
transfer across the incompressible strip, which is difficult to
explain in the framework of one-particle picture. At the same
time, the filling factor $2/3$ is distinguished in this experiment
only by the fact that for the edges of the strip $3/2$, a
complicated structure of collective modes is
expected~\cite{macdonald90,wen91,chamon94}, and interaction
between these modes determines the `excess' equilibration of edge
states. Thus, the experiment ~\cite{fracbutt} for the first time
demonstrated the existence of several branches of collective
excitations at the edge of an FQHE system with the filling factor
$2/3$.

In Ref.~\cite{fraceq06}, the
influence of collective modes on the equilibration at the edge
was studied under the variation of the gate gap width . The study was aimed at transforming a
strongly nonlinear current-voltage characteristic into a
linear one without changing the state of the
two-dimensional electron system in the sample. For this, the
structure of the sample in the quasi-Corbino geometry was
modified: the gate-gap area was made macroscopically large. In this area, an additional
gate was placed. Varying the voltage at the additional gate allows
controlling the width of the interaction area between 10 and 800~$\mu$m.

\begin{figure}
\centerline{\includegraphics[width=\columnwidth]{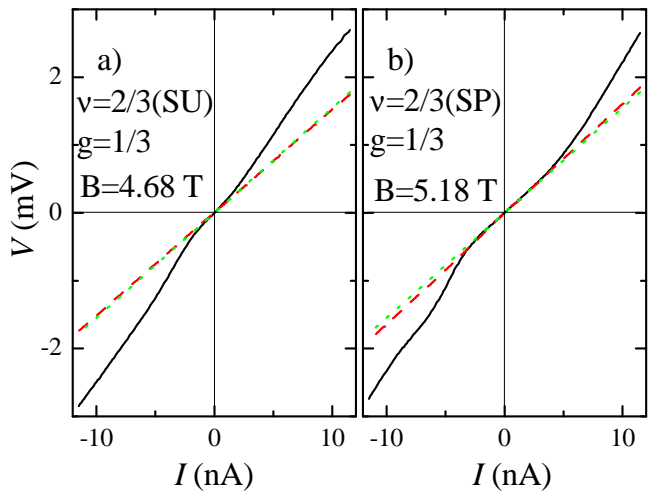}}
\caption{
$I-V$ curves for fractional filling
factors $\nu=2/3, g=1/3$ for narrow (10~$\mu$m, solid line) and
wide (800~$\mu$m, dashed line) interaction regions, for two spin
configurations of $\nu=2/3$: (a) spin unpolarized (SU) state
($B=4.68$~T); (b) spin polarized (SP) state ($B=5.18$~T).
Equilibrium curve (with $R_{eq}=6 h/e^2$) is shown by dots.
(After Ref.~\protect\onlinecite{fraceq06}).
\label{fig12}}
\end{figure}
Transformation of current-voltage characteristics for the filling
factors $\nu=2/3, g=1/3$  is shown in Fig.~\ref{fig12}. As
expected, the $I-V$ curves, initially weakly nonlinear, turn into
linear ones with the slope coinciding with the one found from the
B\"uttiker-Beenakker calculation (\ref{butt_form},\ref{butt_frac}).
The linearity of the central part of the curves means that the
equilibration length does not exceed the interaction area size at
small imbalances. Based on these considerations, the equilibration
length can be estimated to be 10~$\mu$m, which is in agreement
with the results in Refs~\cite{komiyama02,kouwen90,cunning92}
obtained at small imbalances.

\begin{figure}
\centerline{\includegraphics[width=\columnwidth]{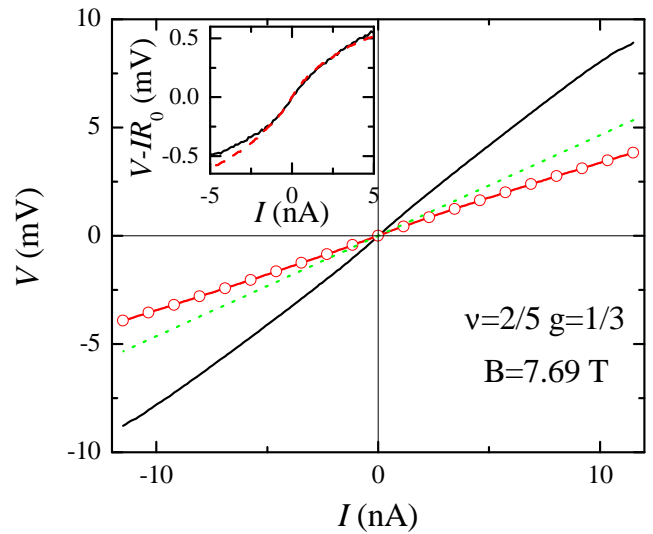}}
\caption{ $I-V$ curves for fractional filling factors $\nu=2/5,
g=1/3$  for narrow (10~$\mu$m, solid line) and wide (800~$\mu$m,
line with open circles) interaction regions. Equilibrium curve
(with $R_{eq}=18 h/e^2$) is shown by dots. Inset shows the
wide-region curve (dash), scaled to the narrow-region one (solid)
in current direction. The linear dependence $IR_0$ with $R_0=28
h/e^2$ is subtracted to highlight the non-linear behavior.
Magnetic field $B$ equals to 7.69~T. (After
Ref.~\protect\onlinecite{fraceq06}) \label{fig13}}
\end{figure}

The most unexpected result is the transformation of $I-V$s
corresponding to the filling factors $\nu=2/5, g=1/3$ (see
Fig.~\ref{fig13}). From the weakly nonlinear $I-V$, which is
situated above the the calculated equilibrium line, the
equilibration length for edge states can be estimated to exceed
10~$\mu$m. As the interaction area increases, the $I-V$ still
remains weakly nonlinear (see the inset in Fig.~\ref{fig13}) but
lies below the equilibrium calculated curve, which would
correspond, in terms of the Beenakker-B\"uttiker single-particle
picture, to \textit{excessive} charge transfer  (by more than a
quarter). Nonlinear curves for both lengths of the interaction
area can be reduced to a single curve by scaling along the current
axis. In this case, the scaling coefficient $q=2.35$ is 40 times
less than the ratio of the interaction area lengths.

We note that prior to Ref.~\cite{fraceq06}, no edge-state
experiments have been carried out for filling factors other than
2/3 and 1/3.   It was predicted~\cite{zhulike04} however, that
collective  modes at the "non-third" edge are expected to have a
considerable impact on the equilibration process. As it was
mentioned above, in Ref.~\cite{fraceq06} at filling factors
$\nu=2/5, g=1/3$ the edge of the 2/5 bulk state has an influence
on the transport effects, because of the small width of the
corresponding compressible strip. Thus, the result~\cite{fraceq06}
is not too unexpected and can be interpreted as the influence of
the collective effects on the equilibration process.

\section{Conclusion}
We summarize the main results of the edge-state investigations in
the IQHE and FQHE regimes:
\begin{itemize}
\item The edge potential of a real system can be considered
smooth in both the IQHE and FQHE regimes. At the edge, there is a
structure of compressible and incompressible strips of electron
liquid.
\item  The B\"uttiker formalism is sensitive only to integral
values, such as the electrochemical potentials of the edges and
total scattering between the edge states.  Therefore, it is not
sensitive to the edge structure and many-body effects. Thus, it is
valid for low imbalances in both the IQHE and FQHE regimes
\item  The structure of edge states for double-layer tunnel-coupled systems corresponds to the structure of the bulk
spectrum and even follows its considerable rearrangements in the
IQHE regime. This leads to a possibility of topological defects
appearing in the edge state structure.
\item Equilibration among the edge states occurs by means of the  electron transport  through incompressible strips.
 This process is fully governed by the single-particle tunnel transparency of the
barrier in such strips in IQHE regime. In contrast, in the FQHE
regime the tunnel density of states has an impact on all effects
related to the transport between edge states, both in the direct
studies of the transport and in the studies of equilibration
between edge states. This many-body tunnel density of states is
governed by the so-called neutral collective excitations at the
edge.
\end{itemize}

\acknowledgments

We wish to thank  V.T.~Dolgopolov for fruitful discussions.
  We gratefully
acknowledge financial support by the RFBR, RAS, the Programme "The
State Support of Leading Scientific Schools", Deutsche
Forschungsgemeinschaft, and SPP "Quantum Hall Systems", under
grant LO 705/1-2. E.V.D. is also supported by MK-4232.2006.2 and
Russian Science Support Foundation.

\end{document}